\documentclass[aps,superscriptaddress,twocolumn,floatfix,nobalancelastpage]{revtex4}
\usepackage{graphicx,siunitx}
\usepackage{amsmath, amssymb}
\usepackage[usenames]{color}
\usepackage{xcolor}
\graphicspath{{Figs/}}

\begin{document}
\title{Liquid migration in shear thickening suspensions flowing through constrictions} 
\author{Rory E. O'Neill}

\affiliation{SUPA and School of Physics and Astronomy, The University of Edinburgh, King's Buildings, Peter Guthrie Tait Road, Edinburgh EH9 3FD, United Kingdom}

\author{John R. Royer}

\affiliation{SUPA and School of Physics and Astronomy, The University of Edinburgh, King's Buildings, Peter Guthrie Tait Road, Edinburgh EH9 3FD, United Kingdom}

\author{Wilson C. K. Poon}
\affiliation{SUPA and School of Physics and Astronomy, The University of Edinburgh, King's Buildings, Peter Guthrie Tait Road, Edinburgh EH9 3FD, United Kingdom}

\date{ \today} 

\begin{abstract}

Dense suspensions often become more dilute as they move downstream through a constriction. We find that as a shear-thickening suspension is extruded through a narrow die and undergoes such liquid migration, the extrudate maintains a steady concentration $\phi_{\rm out}^{\rm LM}$, independent of time or initial concentration. At low volumetric flow rate $Q$, $\phi_{\rm out}^{\rm LM}$ is a universal function of $Q/r_{\rm d}^3$, a characteristic shear rate in the die of radius $r_{\rm d}$, and coincides with the critical input concentration for the onset of LM, $\phi_{\rm in}^{\rm crit}$. We predict this function by coupling the Wyart-Cates model for shear thickening and the `suspension balance model' for solvent permeation through particles. 
\end{abstract}

\maketitle

Suspensions of granular sized particles (radii $a \gtrsim \SI{5}{\micro\meter}$) are ubiquitous in industrial applications, e.g.~molten chocolate \cite{Blanco:2019uo}, ceramic pastes \cite{Carneim:2001ty} and cement~\cite{Banfill:2003aa}. Recent experiments, theory and simulations show that the rheology of suspensions of granular hard particles at high concentration is dominated by the formation of interparticle frictional contacts above some critical `onset stress', $\sigma^*$. Such sliding constraints lead to an increase in viscosity with stress, or shear thickening \cite{Seto:2013aa,  Wyart:2014ve, Guy:2015aa}. 

This new understanding pertains to simple shear, but more complex geometries prevail in applications. Thus, constrictions are frequently encountered, e.g., ceramic paste extrusion through a die or orthopaedic bone cement injection through a syringe. It is unknown to date how recent advances can be applied to these more complex flows, where the material is subjected to significant stress gradients.

Liquid migration (LM)~\cite{Benbow:1993aa}, or self filtration~\cite{Haw:2004aa}, is ubiquitous in flow through a constriction: material becomes more dilute as it moves downstream~ \cite{Maude:1956aa, Seshadri:1968aa}. The solids buildup above the constriction impedes flow, and may lead to jamming. Downstream dilution seriously impacts material strength and stability in ceramics extrusion, and may be fatal in medical applications \cite{ONeill:2017aa}.

\begin{figure}
\center{\includegraphics[width=8.5cm]{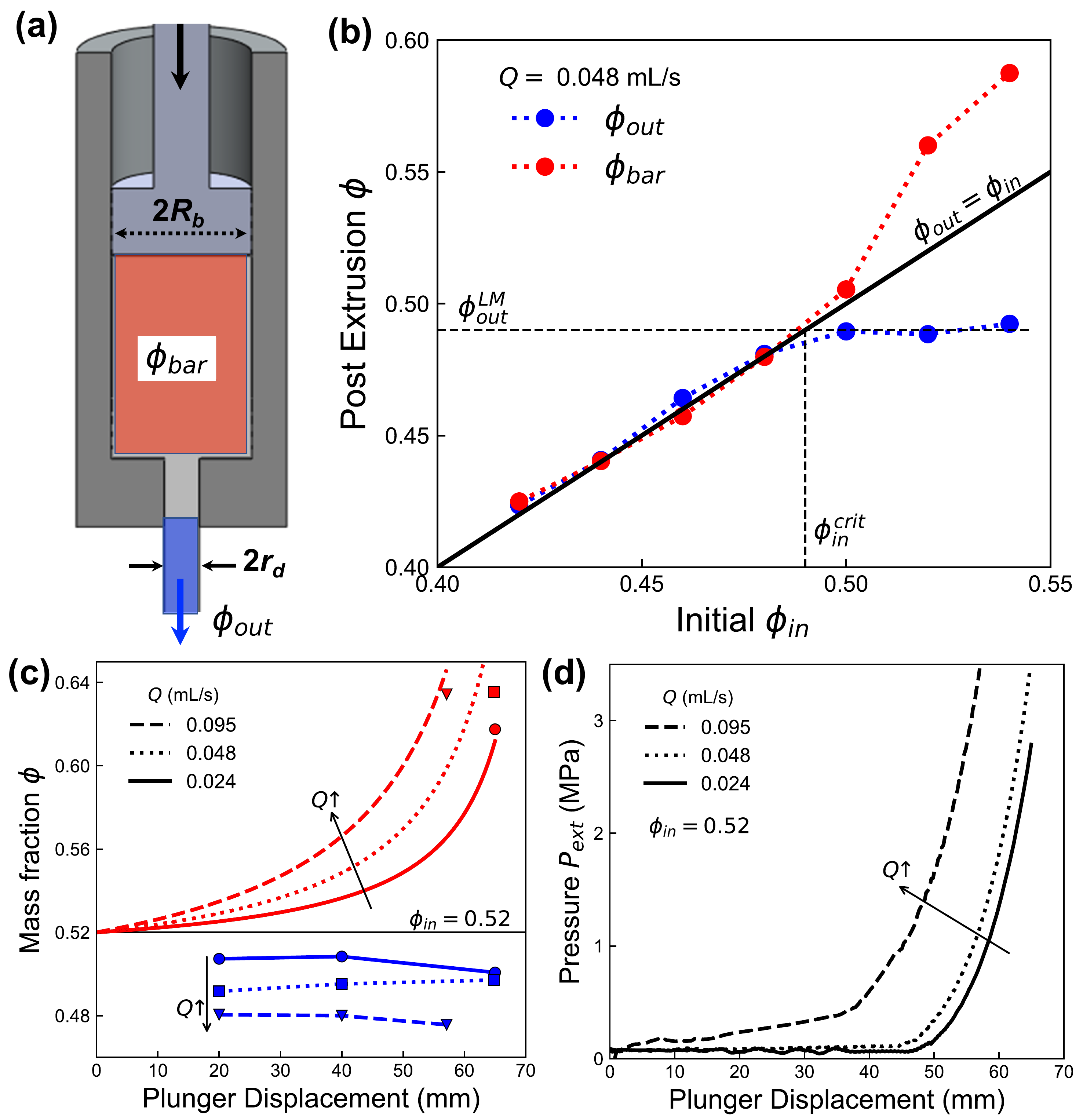}}
\caption{Onset of LM. (a) Extruder schematic. In (b)-(d), $R_{\rm b} =\SI{6.75}{\milli\meter}$ and $r_{\rm d} = \SI{1.7}{\milli\meter}$. (b) Concentration of the extrudate $\phi_{\rm out}$ and material remaining in the barrel $\phi_{\rm bar}$ varying $\phi_{\rm in}$ for $Q= $ 0.048 mL/s. (c)  $\phi_{\rm out}$ at varying intervals of plunger displacements at $\phi_{\rm in} = 0.52$, measured from collected extrudates (blue symbols), and measured post-extrusion $\phi_{\rm bar}$ (red symbols). Red lines: calculated $\phi_{\rm bar}$, assuming constant $\phi_{\rm out}$. Data for $Q = 0.024$ mL/s (circles, solid line), 0.048 mL/s (squares, dotted line) and 0.095 mL/s (diamonds, dashed line).  (d) Extrusion pressure $P_{\rm ext} = F/\pi R_{\rm b}^2$ vs plunger displacement for the same experiments in (c).   }
\label{fig:extrusion_self_filt}
\end{figure}

While many have explored LM in extrusion using specific formulations \cite{Yaras:1994aa, Bayfield:1998aa, Rough:2000aa, Martin:2004aa, Liu:2013aa, ONeill:2015ys}, few have studied the generics using model systems with well-understood rheology to probe the underlying physics \cite{Haw:2004aa, Kulkarni:2010aa}. We investigate LM during die extrusion of cornstarch suspensions, Fig.~\ref{fig:extrusion_self_filt}(a), a model granular shear thickening system \cite{Fall:2008aa, Brown:2012qf, Fall:2015aa, Peters:2016db, Hermes:2016aa, Han:2017aa} (mean particle radii $a\simeq \SI{7}{\micro\meter}$) with a steady-state rheology which fits an analytic model \cite{Wyart:2014ve} for friction-driven thickening \cite{Seto:2013aa,  Guy:2015aa, Comtet:2017aa}. We find that during LM the extruded material (extrudate) maintains a steady solid mass fraction $\phi_{\rm out}^{\rm LM}$, independent of time or initial concentration. Interestingly, $\phi_{\rm out}^{\rm LM}$ is a universal function of $Q/r_{\rm d}^3$ at low to moderate volumetric flow rates $Q$ and all die radii $r_{\rm d}$, which we relate to the cornstarch rheology. 

We extruded cornstarch suspensions at various solid mass fractions $\phi$ (see Supplementary Material for preparation details \cite{supp_mat}) using a custom-built extruder or orthopaedic syringe (OrthoD Group Ltd.) driven by a universal testing machine (Lloyd LS5, AmetekTest), Fig.~\ref{fig:extrusion_self_filt}(a). The custom-built extruder used interchangeable barrels and dies with radii $R_{\rm b}$ and $r_{\rm d}$ respectively, while the orthopaedic syringe had fixed $R_{\rm b}=\SI{6.75}{\milli\meter}$ and $r_{\rm d}=\SI{1.7}{\milli\meter}$. Barrel and die lengths were generally fixed at 40 mm and 10 mm respectively. We drove the plunger at a fixed speed $v_{\rm p}$, giving a volumetric flow rate $Q = \pi R_{\rm b}^2 v_{\rm p}$, and measured the applied force $F$. 

Extrudate was collected in vials and extrusion ceased while material still remained in the barrel, which was recovered by removing the die geometry. The solid mass fraction of the extrudate $\phi_{\rm out}$ and material left in the barrel $\phi_{\rm bar}$ was measured by comparing wet and dry weights~\cite{supp_mat}. 
 
For fixed $\{Q, R_{\rm b}, r_{\rm d}\}$, LM depends on the initial mass fraction of the suspension $\phi_{\rm in}$. With $R_{\rm b} = \SI{6.75}{\milli\meter}$, $r_{\rm d} = \SI{1.7}{\milli\meter}$ and $Q = $ 0.048 mL/s, Fig.~\ref{fig:extrusion_self_filt}(b), $\phi_{\rm in} \simeq \phi_{\rm out} \simeq \phi_{\rm bar}$ below some critical input mass fraction, $\phi_{\rm in} \lesssim \phi_{\rm in}^{\rm crit} \approx 0.49$. When $\phi_{\rm in }$ exceeds $\phi_{\rm in}^{\rm crit}$, $\phi_{\rm out}$ drops below $\phi_{\rm in}$, i.e., LM occurs. We collected a time-lapsed sequence of extrudates from a suspension undergoing LM at $\phi_{\rm in} = 0.52$, Fig.~\ref{fig:extrusion_self_filt}(c). At fixed $Q$, $\phi_{\rm out}$ remains essentially constant at some $\phi_{\rm out}^{\rm LM}$ even as both $\phi_{\rm bar}$, Fig.~\ref{fig:extrusion_self_filt}(c), and the driving pressure, Fig.\ref{fig:extrusion_self_filt}(d), increase dramatically as LM progresses. Moreover, $\phi_{\rm out}^{\rm LM}$ increases with $Q$, Fig.~\ref{fig:extrusion_self_filt}(c). However, once $\phi_{\rm in}$ increases beyond $\phi_{\rm in}^{\rm crit}$ at fixed $Q$, $\phi_{\rm out}$ remains constant at $\phi_{\rm out}^{\rm LM}$, Fig.~\ref{fig:extrusion_self_filt}(a). Non-trivially, to within experimental uncertainties, $\phi_{\rm out}^{\rm LM} = \phi_{\rm in}^{\rm crit}$.

\begin{figure}
\center{\includegraphics[width=8.5cm]{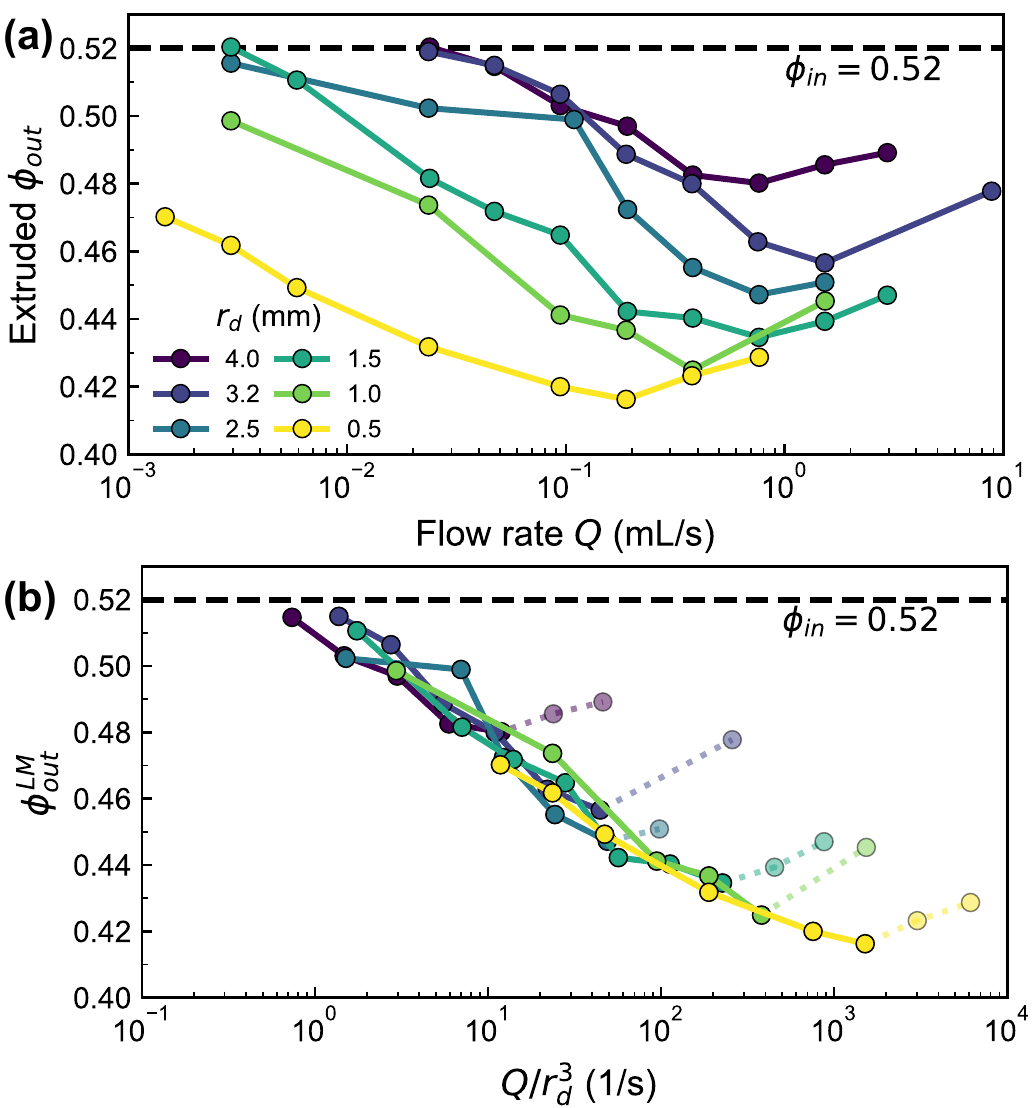}}
\caption{LM under varying flow conditions for $\phi_{\rm in} = 0.52$. (a) Extrudate solid mass fraction, $\phi_{\rm out}(Q)$, for different die radii,~$r_{\rm d}$. (b) Liquid-migrated $\phi^{\rm LM}_{\rm out}$ replotted against the scaling variable $Q/r_{\rm d}^3$. Dotted lines highlight high-$Q$ data beyond the minimum in $\phi_{\rm out}^{\rm LM}$.}
\label{fig:phi_c_Q_r}
\end{figure}

While $\phi_{\rm out}^{\rm LM}$ is independent of compaction within the barrel, it does depend on the flow conditions during extrusion. As $r_{\rm d}$ decreases from \SI{4}{\milli\meter} to \SI{0.5}{\milli\meter} at fixed $\phi_{\rm in} = 0.52$ and $R_{\rm b} = \SI{7.5}{\milli\meter}$, $\phi_{\rm out}^{\rm LM}$ steadily decreases, Fig.~\ref{fig:phi_c_Q_r}(a). That LM increases in smaller constrictions is known \cite{Rough:2000aa}. Note that our smallest $r_{\rm d} \simeq 70a$ is well above the range for arching and clogging in micro-channels \cite{Zuriguel:2014aa} and granular hoppers \cite{To:2001aa, Thomas:2015aa}. As $Q$ increases, $\phi_{\rm out}^{\rm LM}$ decreases to a minimum before increasing. Before the minimum, we find data collapse onto a single master curve, Fig.~\ref{fig:phi_c_Q_r}(b), using the scaling variable $Q/r_{\rm d}^{3}$, which sets the shear rate scale in the die. This collapse breaks down at higher flow rates, suggesting a significant change in extrusion dynamics, so that beyond the minimum LM is no longer solely controlled by shear in the die. Below, we focus on the low-$Q$ regime, where ${\rm d} \phi_{\rm out}^{\rm LM}/{\rm d}Q \leq 0$.

\begin{figure}
\center{\includegraphics[width=8.5cm]{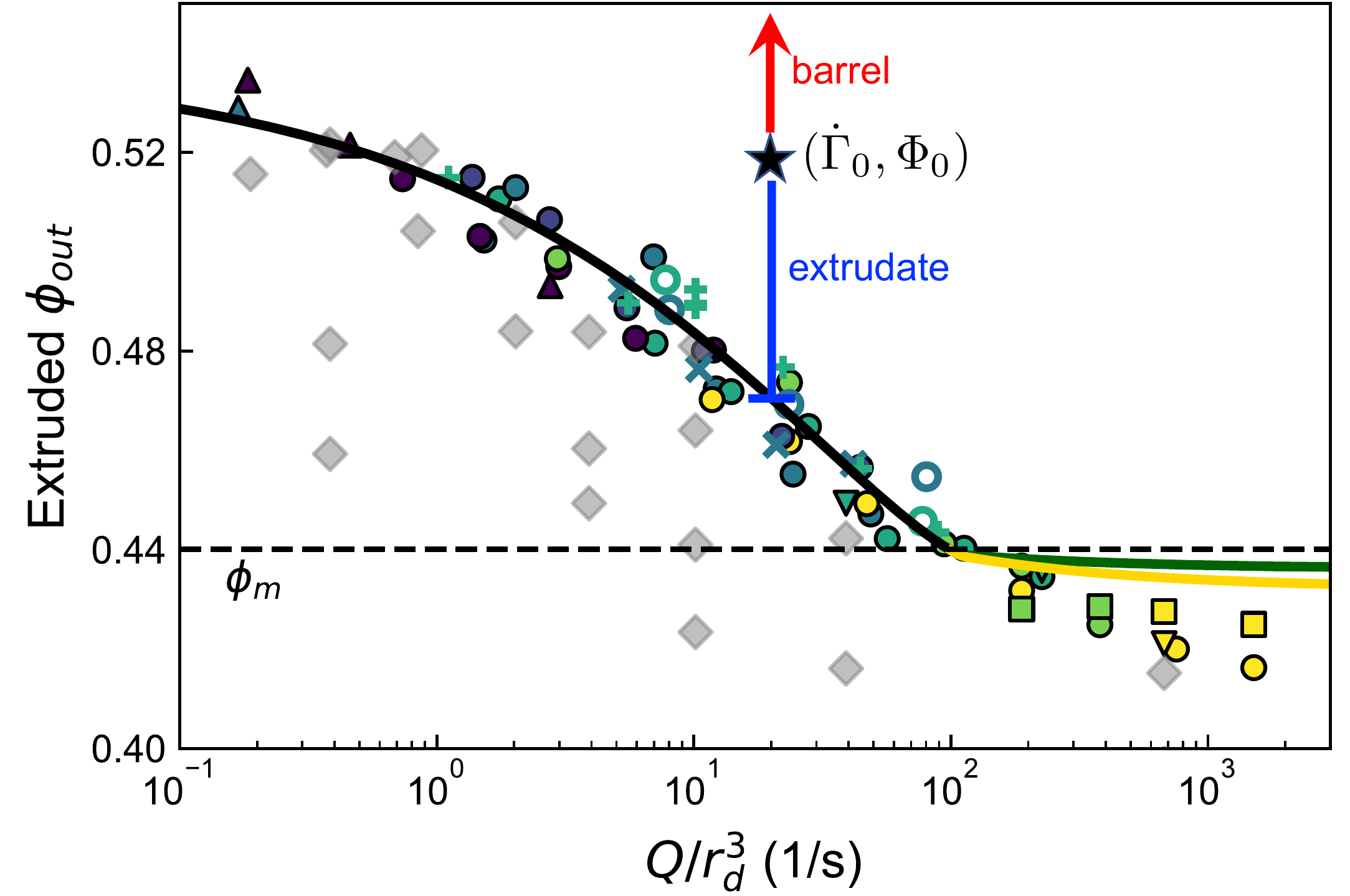}}
\caption{LM state diagram. ({\color{gray} $\blacklozenge$}): $\phi_{\rm out} = \phi_{\rm in}$. Color symbols: $(\phi_c, Q/r_{\rm d}^3)$ for various $(\phi_{\rm in}, R_{\rm b})$. For $R_{\rm b} = \SI{7.5}{\milli\meter}$, circles: $\phi_{\rm in}=0.52$, squares: $\phi_{\rm in} = 0.44$, upside-down triangles: $\phi_{\rm in} = 0.46$  and triangles: $\phi_{\rm in} = 0.54$. For $\phi_{\rm in} = 0.52, $ `$+$' symbols: $R_{\rm b}= \SI{6.75}{\milli\meter}$, open circles: $R_{\rm b}= $ 10 mm and `$\times$' symbols: $R_{\rm b}= $ 12.5 mm. Same $r_{\rm d}$ color scheme as Fig.~\ref{fig:phi_c_Q_r}.  Solid lines: predicted LM phase boundary above $\phi_{\rm m}$. Black: calculated using $\alpha \dot\gamma_ {\rm c}(\phi)$ and using SBM coupled with the WC model for any $r_{\rm d}/a > 70$ (see further Fig.~\ref{fig:explain1}(c) and text discussion); color: calculated using SBM + WC below $\phi_{\rm m}$ for $r_{\rm d}/a = 70$ (yellow) and $r_{\rm d}/a = 140$ (green). }
\label{fig:liq_mig_phase}
\end{figure}

Experiments at different $(\phi_{\rm in}, R_{\rm b})$, Fig.~\ref{fig:liq_mig_phase}, show that all liquid-migrated extrudate concentrations fall on a single master curve, $\phi_{\rm out}^{\rm LM}(Q/r_{\rm d}^3)$. Our data span $\SI{6.75}{\milli\meter} \leq R_{\rm b} \leq  \SI{12.5}{\milli\meter}$,  i.e.~LM is largely unaffected by dynamics within the barrel. Figure~\ref{fig:liq_mig_phase} also displays samples that do not exhibit LM ({$\color{gray} \blacklozenge$}). Their upper bound defines the critical input volume fraction, $\phi_{\rm in}^{\rm crit}$, reinforcing that $\phi_{\rm out}^{\rm LM} = \phi_{\rm in}^{\rm crit}$.

The master curve functions as a `phase boundary'. Below it, LM does not occur. Above it, the boundary gives the extrudate concentration. Thus, a sample above this phase boundary ($\bigstar$, Fig.~\ref{fig:liq_mig_phase}) initially at $\Phi_0$ being extruded at $Q/r_{\rm d}^3 = \dot\Gamma_0$ gives extrudate at a lower concentration given by the intersection of a downward `tie line' from $\bigstar$ to the phase boundary; the other end of this `tie line', giving the barrel concentration, moves towards close packing to satisfy mass conservation. 

The scaling variable $Q/r_{\rm d}^3$ suggests a link between LM and the suspension shear rheology, which we characterized using a rheometer (see Supplementary Material \cite{supp_mat}). Controlling the applied shear stress $\sigma$, we measured the shear rate $\dot{\gamma}$ to obtain $\eta_{\rm r} = (\sigma/\dot{\gamma})/\eta_{\rm s}$, the suspension viscosity relative to that of the solvent ($\eta_{\rm s}$). 

\begin{figure}[t]
\center{\includegraphics[width=8.5cm]{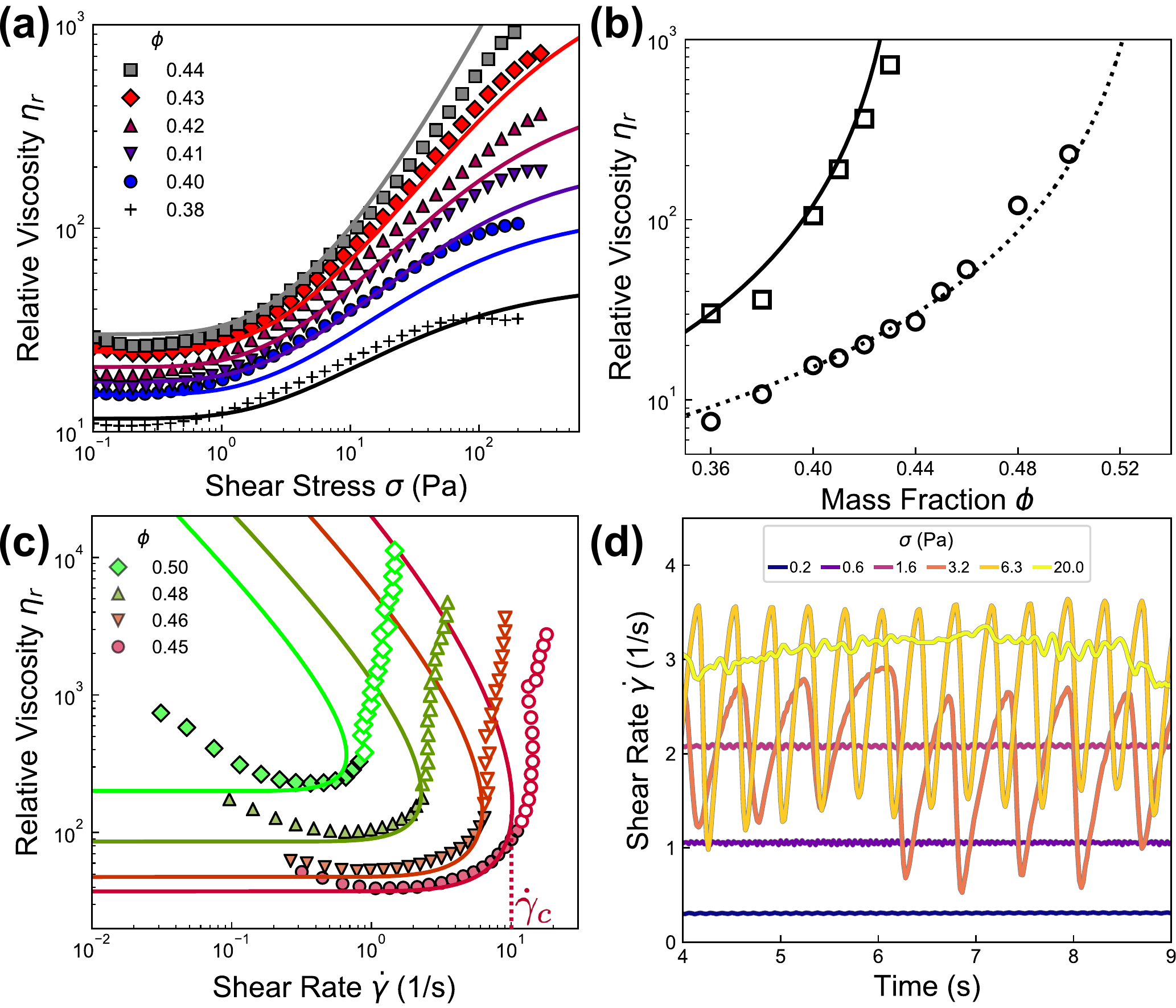}}
\caption{Cornstarch shear rheology. (a) Flow curves $\eta_{\rm r}(\sigma)$ for $\phi < \phi_m$ (points) with WC fits (lines). (b) Low- and high-$\sigma$ viscosities $\eta_1(\phi)$ (circles) and $\eta_2(\phi)$ (squares), respectively, with $\eta_2(\phi)$ taken from the maximum $\eta_{\rm r}(\sigma)$ from flow curves in (a).  Dotted and solid lines: $\eta_{\rm r} = [1-\phi/\phi_{0,m}]^{-2}$, with $\phi_{\rm m}$ determined by fitting the full set of flow curves.  (c) Flow curves for $\phi > \phi_{\rm m}$ (points) with steady (filled symbols) and unsteady (open symbols) flow. Predicted backward-bending flow curves (lines), each with a `nose' at $\dot\gamma_{\rm c}(\phi)$. (d) Onset of fluctuations at $\phi = 0.48$  beyond $\dot \gamma_{\rm c}$. Open symbols in (c) represent a time-average of this unsteady data, and not included in WC fits \cite{unsteady_rheo}. }
\label{fig:rheo}
\end{figure}

Below $\phi = 0.44$, steady-state flow curves $\eta_{\rm r}(\sigma)$ show continuous shear thickening, Fig.~\ref{fig:rheo}(a), and can be described by the Wyart-Cates (WC) model for thickening due to stress-dependent frictional constraints \cite{Wyart:2014ve}. In this model, the viscosity is controlled by two limiting concentrations: $\phi_0$ when all contacts are frictionless, where the low-stress viscosity $\eta_1(\phi)$ diverges, and $\phi_{\rm m}$ when all contacts are frictional, where the high-stress viscosity $\eta_2(\phi)$ diverges. Though edge fracture and interfacial instabilities \cite{Guy:2015aa, Hermes:2016aa} limit full access to $\eta_2(\phi)$, our data is consistent with this picture, Fig.~\ref{fig:rheo}(b). 

Full WC flow curves are given by $\eta_{\rm r}= [1-\phi/\phi_{\rm J}(\sigma)]^{-2}$, where the jamming concentration where $\eta_r$ diverges, $\phi_{\rm J} = \phi_0 [1-f(\sigma)] +  \phi_{\rm m} f(\sigma)$, varies between $\phi_0$ and $\phi_{\rm m}$ as $\sigma$ increases. We model the increasing fraction of frictional contacts  \cite{Wyart:2014ve,Mari:2014aa} by a stretched exponential $f(\sigma) = e^{(-\sigma^*/\sigma)^\beta}$ \cite{Guy:2015aa, Hermes:2016aa}, so that such contacts become above an onset stress $\sigma^*$. We fit our results to this model, extracting $\phi_0$ from $\eta_1(\phi)$ and remaining parameters by simultaneously fitting our full set of flow curves \cite{extrusion_fitsdata}, giving $\phi_0 = 0.538 \pm 0.003$, $\phi_{\rm m} = 0.4401 \pm 0.002$, $\sigma^* = \SI{5.4}{\pascal} \pm \SI{0.7}{\pascal}$ and $\beta = 0.62 \pm 0.03$. To avoid bias towards high viscosities  we fit $\log\eta_{\rm r}$. Though we cannot always reach $\eta_2$ at high $\phi$, the inflection in the flow curves at high-$\sigma$ is sufficient to constrain $\phi_{\rm m}$.

The WC model predicts backward-bending flow curves for $\phi \geq \phi_{\rm m}$, Fig.~\ref{fig:rheo}(c), so above a $\phi$-dependent maximum shear rate $\dot{\gamma}_{\rm c}(\phi)$ the flow curve is no longer defined and steady flow is impossible. In stress-controlled experiments above $\phi_{\rm m}$, we observe a transition from steady to unsteady flow, denoted by changing from filled to open symbols in Fig.~\ref{fig:rheo}(c). In the unsteady regime the suspension viscosity rises sharply, accompanied by large shear rate fluctuations, Figs.~\ref{fig:rheo}(d) \cite{Hermes:2016aa, Bossis:2017aa, unsteady_rheo}.

\begin{figure}
\center{\includegraphics[width=8.5cm]{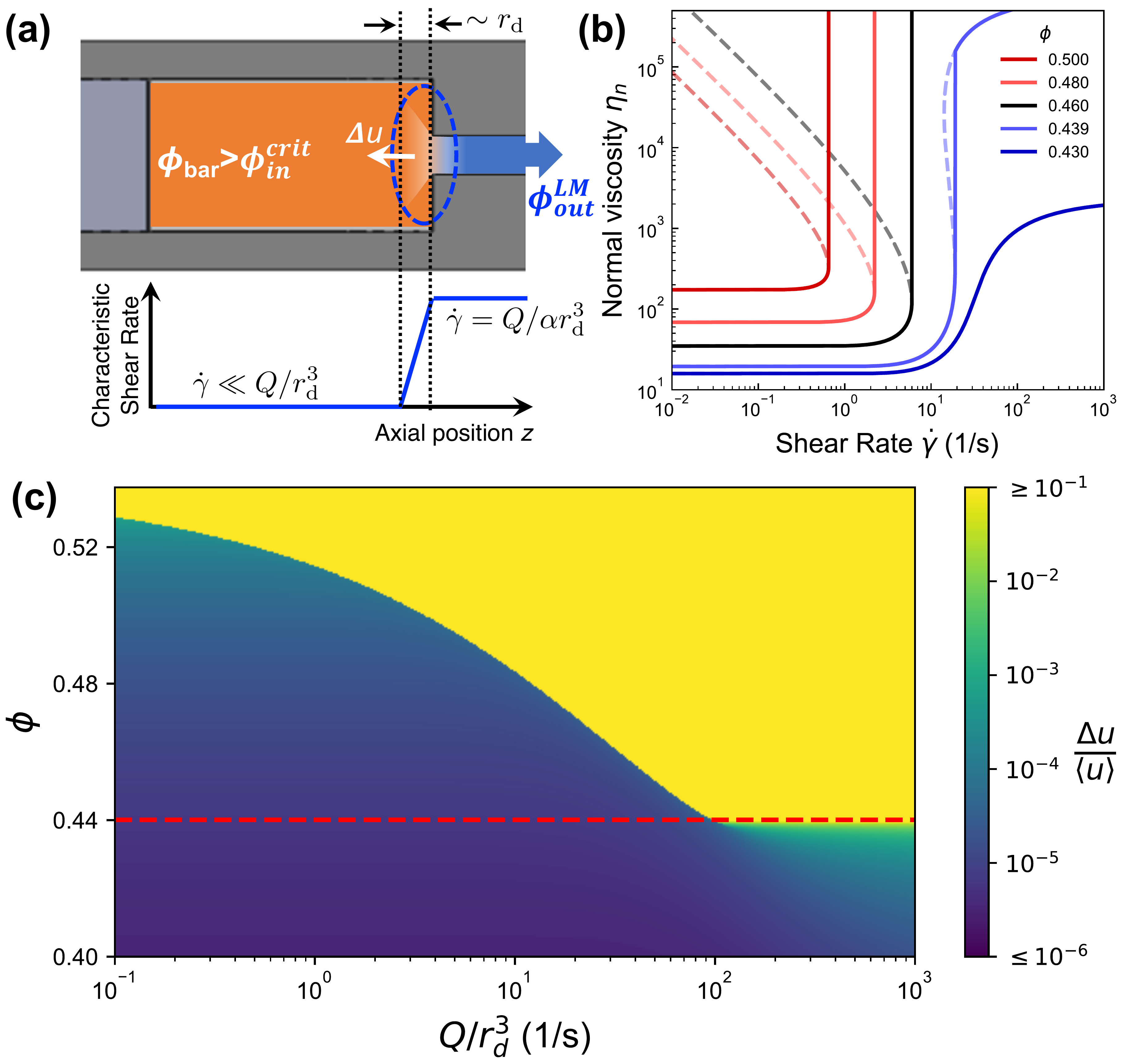}}
\caption{A 1-D model for LM. (a) Schematic of the `dilation zone' above the die. The resulting stress gradient causes a velocity difference $\Delta u = u_{\rm p} - \langle u \rangle$ between the particles and mean flow. (b) Normal viscosity $\eta_{\rm n}(\dot\gamma)$ computed using WC fit parameters.  Dotted lines: full backwards bending or `S' shaped flow curves. Solid lines: profiles used to compute $\Delta \tilde u$ with imposed jumps in the unsteady regime. (c) $\Delta \tilde u(\phi,Q/r_{\rm d}^3)$, computed using $\alpha =4.8$ and $r_{\rm d}/a =140$. Red dashed line: $\phi_{\rm m}$. The contour $\Delta \tilde u(\phi,Q/r_{\rm d}^3) = \epsilon^\dagger=0.004$ gives the SBM+WC phase boundary in Fig. \ref{fig:liq_mig_phase} for $r_{\rm d}/a =140$. }
\label{fig:explain1}
\end{figure}

The quantity $Q/r_{\rm d}^3$ estimates the highest shear rate in the die, which for Newtonian flow occurs at the wall and is given by $4Q/\pi r_{\rm d}^3$. A simple hypothesis is therefore that when $4Q/\pi r_{\rm d}^3$ exceeds the maximum shear rate for stability, $\dot\gamma_{\rm c}(\phi)$, local stresses become arbitrarily large, giving steep stress gradients between die and barrel that would then, as is well known \cite{Leighton:1987aa, Fall:2010aa}, drive migration. 
We in fact find that $Q/r_{\rm d}^3 = \alpha \dot\gamma_c(\phi)$, with $\alpha \approx 4.8$, fits our phase boundary for $\phi \geq \phi_{\rm m}$, Fig.~\ref{fig:liq_mig_phase}(c) (black curve). The large pre-factor $\alpha > 1$ indicates that flow in a finite {\it region} near the die must exceed $\dot\gamma_{\rm c}(\phi)$ to give measurable LM. This region will extend both radially inward and above the die, and since dense suspensions flow with a blunted velocity profile \cite{ Isa:2007aa, Ness:2017aa}, is likely rather thin.

The stress gradients described above will lead to LM both radially, from center to periphery in the die, and axially, from barrel to die, with the latter dominating the dilution of extrudate. We now construct a quasi 1-D model for particle migration along the axis, $z$, using the suspension balance model (SBM) \cite{Nott:1994aa, Morris:1999aa} to quantitatively link migration and local stress gradients. To capture the LM phase boundaries, we neglect time-dependent dynamics and only consider the migration onset starting from an initially uniform suspension. Our {\it ansatz} for $\dot\gamma(z)$ is that it is negligible within the barrel, $\dot\gamma_{\rm b} \ll Qr_{\rm d}^{-3}$, and transitions to some finite $\dot\gamma_{\rm d}$ in the die, Fig.~\ref{fig:explain1}(a), in a zone of size $\sim r_{\rm d}$ immediately upstream to the die. 

During LM, there is a non-zero particle velocity relative to the mean flow, $\Delta u = \frac{\kappa(\phi)a^2} {\eta_{\rm s} }\partial_z \Pi^{\rm p}$, with $\kappa(\phi) = \frac{2}{9}\phi (1-\phi)^4$ the permeability of the particle packing and $\Pi^{\rm p}=  \eta_{\rm s}  \eta_{\rm n}(\phi, \dot\gamma)\dot\gamma$ the shear-induced particle pressure \cite{Nott:1994aa, Morris:1999aa, Deboeuf:2009aa}. The normal viscosity $\eta_{\rm n}$, controlling dissipation due to compressive normal stresses, obeys $\eta_{\rm n}(\phi, \dot\gamma)=[\phi_{\rm J}/\phi(\dot\gamma)-1]^{-2}$ \cite{Boyer:2011nx, Singh:2018sf}, and diverges at $\phi_{\rm J}(\dot\gamma)$, which we take from WC theory.  

We assume that $\dot\gamma$, and so $\Pi^{\rm p}$, are negligible in the barrel outside a small transition zone. In this zone, $ \partial_z \Pi^{\rm p}(\phi,\dot\gamma) \approx \Delta \Pi^{\rm p}/r_{\rm d}\approx \eta_s  \eta_{\rm n}(\phi, \dot\gamma_{\rm d})\dot\gamma_{\rm d}/r_{\rm d} $, where $\dot\gamma_{\rm d}$ is a typical shear rate in the die, Fig.~\ref{fig:explain1}(a). From previous discussion, $\dot\gamma_{\rm d} \sim \dot\gamma_c(\phi) = Q/\alpha r_{\rm d}^3$. Normalizing to the mean flow in the die $\langle u \rangle = Q/\pi r_{\rm d}^2$, we obtain a dimensionless migration speed:
\begin{equation}
\Delta \tilde u(\phi, \dot\gamma_{\rm d}) \equiv \frac{\Delta u}{\langle u \rangle}  \approx \frac{\pi \kappa(\phi) a^2}{\alpha r_d^2} \eta_{\rm n}(\phi, \dot\gamma_{\rm d}).
\end{equation}
Like the shear viscosity,  $\eta_{\rm n}$ bends backwards at $\dot{\gamma}_{\rm c}(\phi)$, Fig.~\ref{fig:explain1}(b), which manifests as a large, abrupt jump in viscosity in rate-controlled flow. To capture this behavior in fixed-$Qr_{\rm d}^{-3}$ extrusion, we impose such jumps in $\eta_{\rm n}$ for $\dot\gamma_{\rm d} \geq \dot{\gamma}_c(\phi)$ when evaluating $\Delta \tilde u(\dot\gamma,\phi)$, Fig.~\ref{fig:explain1}(b), and similar jumps in the `S'-shaped discontinuous shear thickening flow curves \cite{Wyart:2014ve, Pan:2015ab, Mari:2015aa}.

Figure~\ref{fig:explain1}(c) shows $\Delta \tilde u(\phi, Q/r_{\rm d}^3)$ for $\alpha = 4.8$. Although there is finite migration, $\Delta \tilde u>0$, for all $\phi$ and $Q/r_{\rm d}^3$, the jump from negligible [$\Delta \tilde u \sim \mathcal{O}(10^{-5})$, blue] to strong migration [$\Delta \tilde u \sim \mathcal{O}(10^{-1})$, yellow] is {\it very} sharp for $\phi \geq \phi_{\rm m}$, and is associated with the equally abrupt jump in $\eta_{\rm n}$ when $\dot\gamma_{\rm d} \to \dot\gamma_{\rm c}(\phi) = Q/\alpha r_{\rm d}^3$, i.e., at precisely the LM boundary. Formally, if we define the transition from negligible to significant migration to occur at some threshold, i.e.~when $\Delta \tilde u(\phi_c, \dot\gamma_{\rm d}) \geq \epsilon^\dagger$, we recover the observed phase boundary above $\phi_{\rm m}$ for any $10^{-5} \lesssim \epsilon^\dagger \lesssim 10^{-2}$ independent of $r_{\rm d}$ for $r_{\rm d}/a \geq 70$.  

Below $\phi_{\rm m}$, the WC flow curves approach a limiting high-stress viscosity, so there is no longer a maximum possible shear rate $\dot\gamma_c$. Now, the transition from low to high $\Delta \tilde{u}$ is far less abrupt, Fig.~\ref{fig:explain1}(c) (below red dashed line), and where it can be deemed to occur depends on our choice of the threshold, $\epsilon^\dagger$, and also on  $r_{\rm d}/a$.

To choose $\epsilon^\dagger$, we note that the dilation accompanying particle migration is roughly equivalent to $\Delta u$ contributing an extra volume $\Delta V \propto \Delta u \tau_{\rm d} r_{\rm d}^2$ to material in the transition zone above the die, where particles reside for $\tau_{\rm d} \approx r_{\rm d}/\langle u \rangle$. Thus, $|\Delta \phi/\phi| \approx \Delta V/Q\tau_{\rm d} \sim \Delta u/\langle u \rangle \equiv \Delta \tilde u$. Experiments on suspensions of larger particles ($a \gtrsim \SI{50}{\micro\meter}$) below $\phi_{\rm m}$ \cite{Kulkarni:2010aa} detected LM for $|\Delta \phi/\phi| \approx 4 \times 10^{-3}$, which we take to be our $\epsilon^\dagger$. We plot the LM boundary using $\epsilon^\dagger = 0.004$ for two die radii, $r_{\rm d}/a=70$ and 140, in Fig.~\ref{fig:liq_mig_phase}. The fit to data below $\phi_{\rm m}$ is credible, but poorer than for $\phi > \phi_{\rm m}$. Using smaller $\epsilon^\dagger$ produces better fits, but is hard to motivate physically. Whatever the choice of $\epsilon^\dagger$ and $r_{\rm d}/a$, our model captures the abrupt change in slope of the LM boundary at $\phi_{\rm m}$. 

Summarizing, we have characterized LM during extrusion of shear-thickening cornstarch suspensions. The onset concentration at low to moderate flow rates lies on a universal boundary if data for different flow rates and die radii are plotted against $Qr_{\rm d}^{-3}$, an estimate of the maximum shear rate in the die. The locus where $Qr_{\rm d}^{-3} \to \alpha \dot\gamma_{c}(\phi)$ fits well the observed LM boundary above $\phi_{\rm m}$ with $\alpha \approx 4.8$, suggesting that flow in a finite region near the die entrance must be unstable for  appreciable LM. The instability point $\dot\gamma_{c}(\phi)$ can be estimated by where $\eta_{\rm r}$ dramatically increases in controlled-stress rheology, Fig.~\ref{fig:rheo}(c). We therefore have a theory for LM at $\phi \geq \phi_{\rm m}$ up to a single dimensionless parameter independent of any other theoretical model. It is also possible to obtain $\dot\gamma_{c}(\phi)$ by fitting bulk rheology data to WC theory. Coupling this to the SBM for particle migration in a simple 1D model gives a semi-quantitative prediction of the boundary below $\phi_{\rm m}$. A more sophisticated theory of LM accounting for radial migration \cite{Hampton:1997aa, Frank:2003aa, Isa:2007aa, Oh:2015aa} and extensional flow \cite{Seto:2017aa, Cheal:2018aa} may obviate the need for $\alpha$ and produce better agreement at $\phi < \phi_{\rm m}$.

Previously, LM had been modeled using finite-time-step methods to simulate extrusion, and either a 1-D \cite{Rough:2002aa, Khelifi:2013aa} or 2-D finite-element model \cite{Patel:2007aa, Patel:2017aa} to describe the paste. These empirical paste models rely on material parameters not directly extractable from shear rheology. Particle-based simulations of extrusion, which lack an explicit fluid phase, reproduce localized shear and stress gradients near the die entry but not LM, highlighting the importance of such gradient-driven flows \cite{Ness:2017aa}.  A model for LM in suspensions of larger granular particles based on suspension balance exists \cite{Kulkarni:2010aa}, but requires measured particle pressures as input. Our model uses bulk rheology data to capture LM.

While we find increasing LM with increasing (low to moderate) flow rate in shear thickening suspensions, LM increases with decreasing flow rates in many other pastes \cite{Bayfield:1998aa, Rough:2000aa, Liu:2013aa, ONeill:2017aa}, possibly because attractive or adhesive interaction between particles gives rise to yield-stress and shear-thinning behavior. A recently-proposed constraints-based extension of the WC model to include such interactions \cite{Guy:2018aa} may allow application of our approach here to a broader range of pastes. 

\begin{acknowledgments}
Funding came from the EPSRC (EP/N025318/1) and the NSF under Grant No.~NSF PHY-1748958 through
the KITP program on the Physics of Dense Suspensions.
\end{acknowledgments}

\bibliography{extrusion_library_simple}

\begin{thebibliography}{54}
\bibitem[{ext()}]{extrusion_fitsdata}
\bibinfo{note}{The data in this article, including fitting routines written in
  Python + Scipy, are available online at ::insert url:: .}
\bibitem[{uns()}]{unsteady_rheo}
\bibinfo{note}{In the unsteady flow regime, the time-averaged viscosity should
  not be expected to faithfully reflect the underlying bulk rheology. However,
  the instability onset and sharp rise in viscosity at $\dot\gamma_{\rm c}$ is
  clearly evident in this data.}
\end{thebibliography}
\bibliographystyle{apsrev}

\end{document}